\documentclass[pra,twocolumn,floatfix,showpacs]{revtex4}%
\usepackage[english]{babel}
\usepackage{amsmath}
\usepackage{color}
\usepackage{epsfig}
\usepackage{graphicx}
\usepackage{bm}

\usepackage{times,amsmath,amssymb}
\usepackage{amsfonts}
\usepackage{amssymb}
\usepackage{color}
\definecolor{myred}{RGB}{168,5,14}

\newcommand{\tr}{\mathop{\mathrm{tr}}\limits}
\newcommand{\trb}{\mathop{\mathrm{tr}_{1,N}}\limits}

\begin{document}
\title{Obtaining pure steady states in nonequilibrium quantum systems
  with strong dissipative couplings } \author{Vladislav Popkov}
\thanks{E-mail: popkov@uni-bonn.de} \affiliation{Institut f\"{u}r
  Teoretische Physik, Universit\"{a}t zu K\"{o}ln, Z\"ulpicher
  Strasse 77, K\"{o}ln, Germany.}  \affiliation{HISKP,
  University of Bonn, Nu\ss allee 14-16, 53115 Bonn, Germany.}
\affiliation{ Centro Interdipartimentale per lo studio di Dinamiche
  Complesse, Universit\`a di Firenze, via G.  Sansone 1, 50019 Sesto
  Fiorentino, Italy }

\author{Carlo Presilla} \affiliation{Dipartimento di Fisica, Sapienza
  Universit\`a di Roma, Piazzale Aldo Moro 2, Roma 00185, Italy}
\affiliation{Istituto Nazionale di Fisica Nucleare, Sezione di Roma 1,
  Roma 00185, Italy}

\begin{abstract}
  Dissipative preparation of a pure steady state usually involves a
  commutative action of a coherent and a dissipative dynamics on the
  target state.  Namely, the target pure state is an eigenstate of
  both the coherent and dissipative parts of the dynamics. We show
  that working in the Zeno regime, i.e. for infinitely large dissipative
  coupling, one can generate a pure state by a non commutative action,
  in the above sense, of the coherent and dissipative dynamics. A
  corresponding Zeno regime pureness criterion is derived.  We
  illustrate the approach, looking at both its theoretical and
  applicative aspects, in the example case of an open $XXZ$ spin-$1/2$
  chain, driven out of equilibrium by boundary reservoirs targeting
  different spin orientations.  Using our criterion, we find two
  families of pure nonequilibrium steady states, in the Zeno regime,
  and calculate the dissipative strengths effectively needed to
  generate steady states which are almost indistinguishable from the
  target pure states.
\end{abstract}

\date{\today }

\maketitle

\section{Introduction}
One indispensable pre-requisite for quantum information processing is
preparing a given quantum state and maintaining it for a sufficiently
long time.  A promising perspective in generating quantum states with
desired properties is offered by using a controlled dissipation.
Instead of producing a detrimental decoherent effect on the quantum
system, the controlled dissipation can help to create and preserve the
coherence.  With the help of the controlled dissipation, one can
prepare and maintain entangled qubit
states~\cite{Knight2000,Kastoryano2011PRL,
  LinNature2013Bell,ShankarNature2013Bell,Kienzler2015Science,TicozziViola2012},
perform universal quantum computational
operations~\cite{CiracNature2009,BlatterPhysReports08,BlattNature2013},
generate and replicate entanglement between macroscopic
systems~\cite{Cirac2011PRL107,Stannigel2012NJPcascades,Adesso}, and store
and protect quantum memory~\cite{CiracPRAmemory}.  Dissipative state
engineering methods are robust, since, due to the dissipative nature
of the process, the system is driven towards its nonequilibrium steady
state (NESS) independently of the initial state and of the presence of
perturbations.

Dissipative pure state engineering typically requires
\textit{commutative} actions on the target state by both the coherent
and dissipative parts of the effective
dynamics~\cite{Cirac1993forKinzler,Kienzler2015Science,TicozziViola2012,
  Yamamoto05,ZollerPRA08,KrausNature2008,Cormick2013NJP}.  In other
words, the target state is required to be an eigenstate of the
Hamiltonian and of all quantum jump operators, see
Eq.~(\ref{PureNESSsuffientCondition})\cite{YamamotoRemark}.  On the
other hand, generic non-commutative coherent and dissipative actions
result in a mixed steady state~\cite{Diehl2010DynamicPhaseTransition}.

In this paper, we demonstrate that by applying sufficiently strong
dissipative couplings, one can generate steady states, which are
arbitrarily close to pure states, for \textit{non-commutative}
dissipative and coherent dynamics. Namely, while the target pure state
is still required to be an eigenstate with respect to the quantum jump
operators, it is not generically an eigenstate of the
Hamiltonian. This is not in conflict with previous results, since the
exact pure NESS is attained only in the Zeno limit, i.e. in the limit
of infinitely strong dissipative action, where the NESS pureness
criteria~\cite{Yamamoto05,ZollerPRA08} are not valid.

The Zeno regime belongs nowadays to a standard toolbox of dissipative
protocols~\cite{ZenoStaticsExperimentalReview}.  It is usually
associated with an effect of freezing the whole quantum system, or
freezing some degrees of freedom and accelerating some others (static
Zeno effect, dynamic Zeno effect, anti-Zeno
effect)~\cite{ZenoStaticsExperimental,ZenoDynamicsExperimental,Zeno}.
In the following, we derive a general criterion of steady-state
pureness which applies exactly in the Zeno regime but can be used to
generate an almost pure NESS for sufficiently strong dissipative
couplings.  We demonstrate the applicability of our criterion by
obtaining two classes of pure stationary states in nonequilibrium
boundary-driven Heisenberg $XXZ$ spin chains, both in the critical and
noncritical phases.  Moreover, we show that, in practice, reaching the
Zeno regime is not necessary, since applying a dissipation above a
finite strength is sufficient to obtain pure steady states with
arbitrary pre-set pureness.

\section{Zeno regime pure NESS criterion}
We consider an open quantum system in contact with an external
environment.  The effective time evolution of the reduced density
matrix $\rho$ of the system is described by a quantum master equation
in the Lindblad form \cite{Petruccione,PlenioJumps,ClarkPriorMPA2010},
\begin{align}
  \frac{\partial\rho}{\partial t}= -i\left[ H,\rho\right] +\Gamma
  \mathcal{D}[\rho],
  \label{LME}%
\end{align}
where $H$ is the Hamiltonian representing the coherent part of the
evolution, $\Gamma$ measures the strength of the dissipative coupling,
and $\mathcal{D}[\rho]$ is the Lindblad dissipator,
\begin{align}
  \mathcal{D}[\rho]= \sum_{\alpha} \left( L_\alpha \rho L^\dag_\alpha-
    \frac{1}{2} \left( L^\dag_\alpha L_\alpha \rho + \rho
      L^\dag_\alpha L_\alpha \right) \right),
  \label{D[ro]}%
\end{align}
defined in terms of a set of Lindblad, or quantum jump, operators,
$\{L_\alpha\}$.  We set $\hbar=1$ and $J=1$, where $J$ is a global
energy factor which multiplies $H$, measuring energy in units of $J$,
time in units of $\hbar/J$, and $\Gamma$ in units of $J/\hbar$.  A
NESS is a fixed point solution of the dynamical Lindblad
equation~(\ref{LME}).  We shall assume that the NESS is unique.  It is
easy to see that the NESS is a pure state, namely,
$\rho_\mathrm{NESS}(\Gamma)= | \Psi \rangle \langle \Psi |$, if $|
\Psi \rangle$ is an eigenstate of the Hamiltonian and a dark state
(i.e., an eigenstate with zero eigenvalue) with respect to all
Lindblad operators,
\begin{align}
  H | \Psi \rangle= \lambda | \Psi \rangle \quad\mbox{and}\quad
  L_\alpha | \Psi \rangle =0, \quad\mbox{for all $\alpha$}.
  \label{PureNESSsuffientCondition}%
\end{align}
Most theoretical studies and experimental protocols rely on this
sufficient condition (\ref{PureNESSsuffientCondition}) for
dissipatively preparing pure states.  It often happens, however, that
for the given set $H,\{L_\alpha\}$, no pure state satisfying the
conditions (\ref{PureNESSsuffientCondition})~\cite{YamamotoRemark} can
be found.  In those cases, it is worth formulating a less demanding
criterion by requiring $\rho_\mathrm{NESS}(\Gamma)$ to become pure
only in the Zeno limit $\Gamma \rightarrow \infty$. We then assume
that for sufficiently large $\Gamma$, the following expansion in
powers of $(1/\Gamma)^{k}$ exists:
\begin{align}
  \rho_\mathrm{NESS}(\Gamma) = | \Psi \rangle \langle \Psi |+
  \sum_{k=1}^{\infty}\Gamma^{-k}\rho^{(k)}, \label{PT_largeCouplings}%
\end{align}
where the first term of the expansion $\rho^{(0)}= | \Psi \rangle
\langle \Psi |$ is a pure state.  Inserting the time-independent state
(\ref{PT_largeCouplings}) into Eq.~(\ref{LME}) and comparing the terms
at different orders of $\Gamma$, we obtain
\begin{align}
  \mathcal{D}[| \Psi \rangle \langle \Psi |] =0
  \label{D[ro0]=0}%
\end{align}
and the recurrence relations
\begin{align}
  i[H,\rho^{(k)}]= \mathcal{D}[\rho^{(k+1)}], \qquad k=0,1,2, \dots,
  \label{Recurrence0}%
\end{align}
which have the formal solution
\begin{align}
  \rho^{(k+1)}= \mathcal{D}^{-1}[i[H,\rho^{(k)}]], \qquad k=0,1,2,
  \dots.
  \label{Recurrence}%
\end{align}

The existence of $\mathcal{D}^{-1}[i[H,\rho^{(k)}]]$ is granted if and
only if $[H,\rho^{(k)}]$ lies entirely in the subspace orthogonal to
the kernel of $\mathcal{D}$, i.e.
\begin{align}
  P_{\ker \mathcal{D} } ([H,\rho^{(k)}]) = 0, \hspace{1cm} k=0,1,2...,
  \label{SecularConditions}%
\end{align}
where $P_{\Omega}$ denotes the orthogonal projector on $\Omega$. In
particular, the zeroth-order condition reads
\begin{align}
  P_{\ker \mathcal{D} } ([H,| \Psi \rangle \langle \Psi |]) = 0.
  \label{SecularConditions0}%
\end{align}
Conditions (\ref{D[ro0]=0}) and (\ref{SecularConditions}), which, for
brevity, will be named the Zeno regime pure NESS criterion, substitute
the criterion (\ref{PureNESSsuffientCondition}) in the limit
$\Gamma\to\infty$.  As we will demonstrate in the following, the Zeno
regime pure NESS criterion is less restrictive than criterion
(\ref{PureNESSsuffientCondition})\cite{YamamotoRemark}.  Moreover,
satisfying Eq.~(\ref{D[ro0]=0}) and just the zeroth order necessary
condition (\ref{SecularConditions0}) can be enough to find a pure NESS
in the Zeno limit.  By continuity, for sufficiently large dissipative
coupling $\Gamma$, the actual NESS will be arbitrarily close to the
pure state.  The target state $| \Psi \rangle$ is not an eigenstate of
the Hamiltonian $H$; otherwise the condition
(\ref{SecularConditions0}) becomes trivial. On the other hand, the
condition (\ref{D[ro0]=0}) implies \cite{Yamamoto05,ZollerPRA08} that
the target state is an eigenstate of the quantum jump operators
$\{L_\alpha\}$. Thus, the actions of the coherent and dissipative
parts of the dynamics on $| \Psi \rangle$ are
\textit{non-commutative}, $H L_\alpha | \Psi \rangle \neq L_\alpha H |
\Psi \rangle$, which implies that the target pure state cannot be
\emph{exactly} prepared for any finite $\Gamma$.

\section{Heisenberg spin chains}
To test the Zeno regime pure NESS criterion, we consider an open $XXZ$
Heisenberg spin chain with Hamiltonian
\begin{align}
  H=\frac{1}{2} \sum_{j=1}^{N-1}
  \left(  \sigma_{j}^{x}\sigma_{j+1}^{x}+ \sigma_{j}^{y}\sigma_{j+1}^{y}+
    \Delta \left( \sigma_{j}^{z}\sigma_{j+1}^{z}- I \right) \right),
  \label{Hamiltonian}%
\end{align}
where $\Delta$ is the dimensionless anisotropy parameter measuring the
ratio between the couplings of the $Z$ and $XY$ spin components, and a
dissipator with just two Lindblad operators, $L_1=\mathcal{L}_{L}$ and
$L_2=\mathcal{L}_{R}$ acting locally on the ``left'' and ``right''
boundary spins only.  The operators $\mathcal{L}_L$ and
$\mathcal{L}_R$ favor an alignment of the boundary spins at $k=1$ and
$k=N$ along the vectors $\vec{l}_{L},\vec{l}_{R}$ defined by
longitudinal and azimuthal coordinates as
\begin{align*}
  \begin{array}{l}
    \vec{l}_{L}= (\sin\theta_{L}\cos\varphi_{L},\sin\theta_{L}\sin
    \varphi_{L},\cos\theta_{L}),
    \\
    \vec{l}_{R}=(\sin\theta_{R}\cos\varphi_{R},\sin\theta_{R}\sin
    \varphi_{R},\cos\theta_{R}).
  \end{array}
\end{align*}
If $\vec{l_L}\neq \vec{l}_R$, then there is a boundary gradient
leading to a NESS with nonzero current.  For specific boundary
gradients, the NESS of this model has been calculated analytically at
arbitrary dissipation strength
\cite{ProsenExact2011,MPA2013-1,MPA2013-2,2015ProsenReview}.  The
explicit form of $\mathcal{L}_{L},\mathcal{L}_{R}$ is given in
Appendix \ref{SM1}, where we also detail the content of
Eq.~(\ref{SecularConditions}) and the calculation of the
super-operator inverse $\mathcal{D}^{-1}$.  In the Zeno limit, the
boundary spins $1,N$ are projected into the states described by the
\emph{one-site} density matrices
\begin{align}
  \rho_{L} & =\frac{1}{2}\left(
    I+\vec{l}_{L}\cdot\vec{\sigma}_{1}\right),
  \label{RoL}\\%
  \rho_{R} & =\frac{1}{2}\left(
    I+\vec{l}_{R}\cdot\vec{\sigma}_{N}\right) .
  \label{RoR}%
\end{align}
These are single qubit pure states, $\tr \rho_{L}^2=\tr \rho_{R}^2=1$.

We look for a zeroth-order pure NESS, $\rho_\mathrm{NESS} = | \Psi
\rangle \langle \Psi |$, in the factorized form
\begin{align}
  | \Psi \rangle = | \psi_1 \rangle \otimes | \psi_2 \rangle \otimes
  \dots | \psi_{N} \rangle, 
\label{PsiFactorized}%
\end{align}
with $| \psi_k \rangle$ satisfying the generalized divergence
condition
\begin{align}
  h \left(|\psi_k \rangle \otimes |\psi_{k+1}\right \rangle) =&\ \mu_k
  |\psi_k\rangle \otimes |\psi_{k+1}\rangle \nonumber\\ &+ |U_k\rangle
  \otimes |\psi_{k+1}\rangle - |\psi_{k}\rangle \otimes |U_k\rangle,
  \label{GeneralizedDivergenceCondition}%
\end{align}
where $h$ is a local density of the Hamiltonian~(\ref{Hamiltonian})
and $|U_k\rangle$ is a local unknown vector.  Substituting expression
(\ref{GeneralizedDivergenceCondition}) into
Eq.~(\ref{SecularConditions0}), we find that this is satisfied if and
only if
\begin{subequations}
  \begin{align}
    &\sum_{k=1}^{N-1} \left(\mu_k-\mu_k^* \right) = 0,
    \\
    &R_k = \tilde{R}_{k},\qquad k=1,\dots,N,
    \\
    &\tr R_1 =\tr \tilde{R}_N= 0,
  \end{align}
  \label{ZerothOrderConditionsForR}%
\end{subequations}
where $R_k = | U_k \rangle \langle \psi_k | - | \psi_k \rangle \langle
U_k |$ and $\tilde{ R}_k = | U_{k-1} \rangle \langle \psi_k | - |
\psi_k \rangle \langle U_{k-1} |$.

\textit{Proof.}  Denoting $\rho_k=|\psi_{k}\rangle \langle \psi_{k} |$
and using Eq.~(\ref{GeneralizedDivergenceCondition}), we rewrite the
commutator $[H,\rho_\mathrm{NESS}]$ as
\begin{align}
  [H,\rho_\mathrm{NESS}] &= \sum_{k=2}^{N-1} \rho_1 \otimes \dots
  \otimes (R_k - \tilde{R}_{k}) \otimes \cdots \otimes \rho_{N}
  \nonumber\\
  &+ R_1 \otimes \rho_2 \otimes \cdots \otimes \rho_{N} - \rho_1
  \otimes \cdots \otimes \rho_{N-1} \otimes \tilde{R}_{N}
  \nonumber\\
  &+ \sum_{k=1}^{N-1} \left(\mu_k -\mu_k^*\right) \rho_\mathrm{NESS}.
  \label{CommutatorWithH}%
\end{align}
Requiring that Eq.~(\ref{SecularConditions0}) is satisfied and taking
into account that $\tr(\rho_1)=\tr(\rho_N)=1$ and $\tr(A\otimes
B)=\tr(A)\tr(B)$, we obtain (\ref{ZerothOrderConditionsForR}).

The criterion (\ref{PureNESSsuffientCondition}) would not, in the
present example, provide any nontrivial solution: the NESS is not pure
for any finite $\Gamma$ and for any boundary polarization gradient.
The only solution of Eqs.~(\ref{PureNESSsuffientCondition}) is
obtained for identical boundary conditions, $\vec{l_L}= \vec{l}_R$,
and fixed anisotropy, $\Delta=1$, and it corresponds to a trivial
ferromagnetic state $\rho=(\rho_{L})^{\otimes_N}$.  Conversely, using
the Zeno regime pure NESS criterion, we readily find the following two
nontrivial families of solutions.

\subsection{Boundary twisting in the $XY$ plane}  
Let us choose the boundary polarizations in the $XY$ plane.  Due to
isotropy, this choice can be parametrized by a single angle $\Phi$
between the left and right boundary polarizations, i.e., we can put
${\vec l}_L = (1,0,0)$, ${\vec l}_R = (\cos \Phi,\sin \Phi,0)$.
Various properties of the $XXZ$ model with boundary twisting in the
$XY$ plane for strong and weak driving have been investigated for
$\Phi=\pi/2$ and arbitrary $\Delta$ in
\cite{2012XYtwist,2013XYWeakDriving}, while for the isotropic case
$\Delta=1$, the full analytic NESS for arbitrary $\Gamma,\Phi$ has been
obtained in \cite{MPA2013-1,MPA2013-2}.

\begin{figure}[ptbh]
  \begin{center}
    \includegraphics[width=0.3\columnwidth]{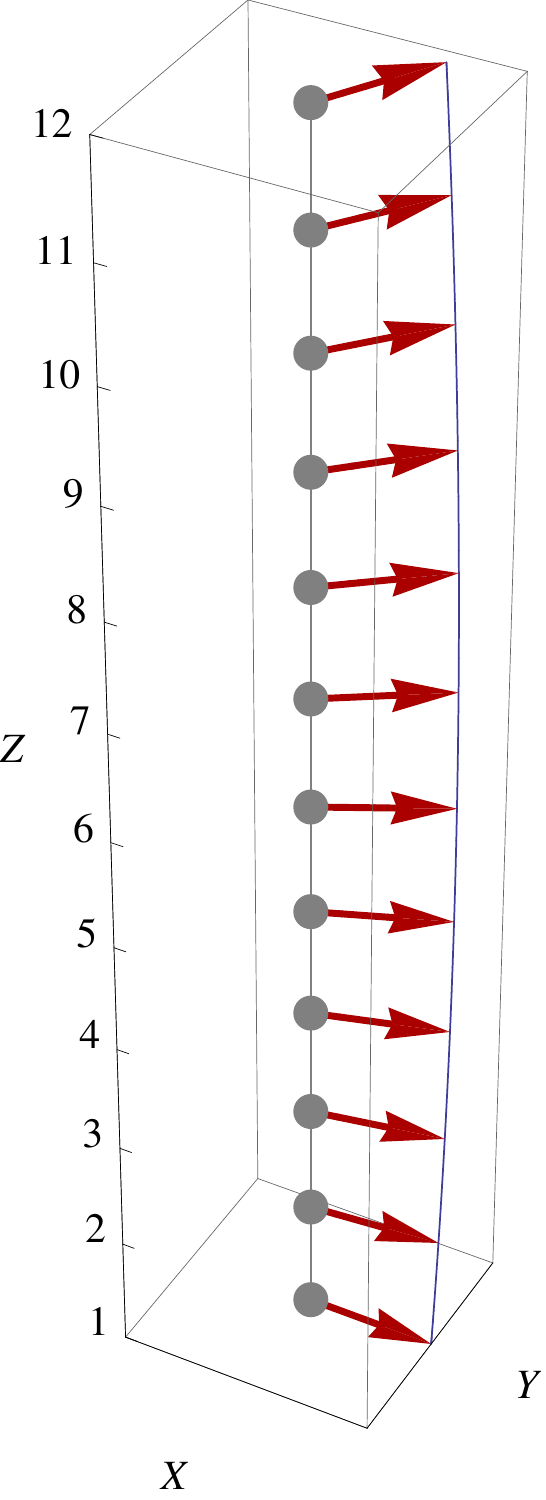}
    \includegraphics[width=0.3\columnwidth]{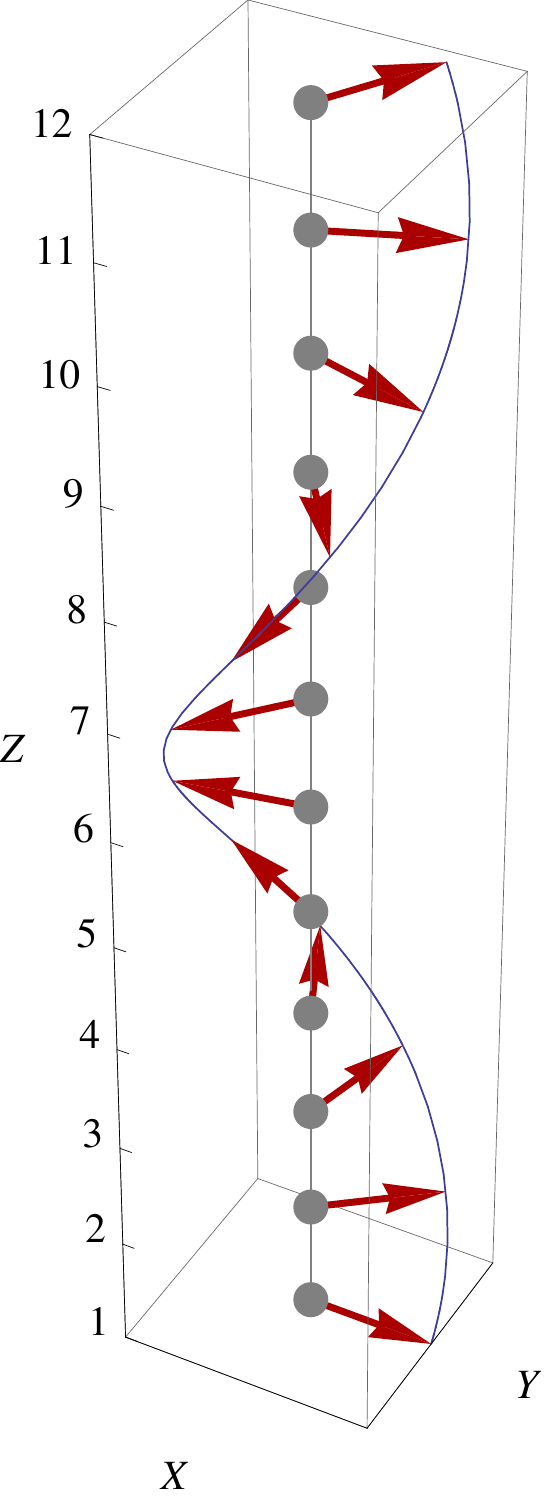}
    \includegraphics[width=0.3\columnwidth]{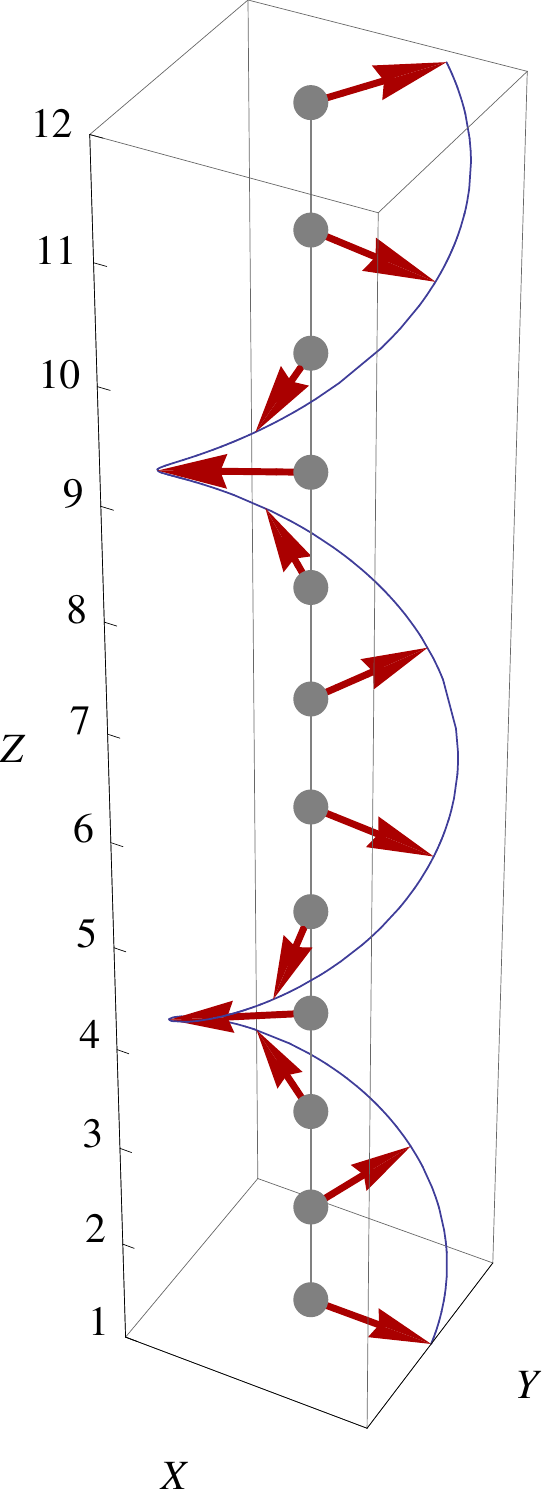}
  \end{center}
  \caption{ (Color online) Orientation, for a pure NESS in the Zeno
    regime, of the lattice spins of a driven $XXZ$ chain arranged
    along the $Z$ axis with boundary twisting in the $XY$ plane for
    winding number $m=0,1,2$ (left to right).  The $Z$ labels indicate
    the lattice points along the chain.  Parameters: $N=12$,
    $\Phi=\pi/3$.}
  \label{FigSpinsXY}%
\end{figure}

We look for a solution of Eq.~(\ref{GeneralizedDivergenceCondition})
taking $| \psi_k \rangle$ in the form
\begin{align}
  | \psi_k \rangle= \frac{1}{\sqrt{2}} \left(
    \begin{array}{c}
      e^{-i\frac{\varphi_k}{2}}
      \\
      e^{i\frac{\varphi_k}{2}}
    \end{array}
  \right),
\end{align}
which corresponds to a local spin polarization ${\vec l}_k= (\cos
\varphi_k,\sin \varphi_k,0)$.  As detailed in Appendix \ref{SM2}, such a
solution exists and, via Eq.~(\ref{ZerothOrderConditionsForR}), is
also a solution of Eq.~(\ref{SecularConditions0}), provided that
$\varphi_{k+1} - \varphi_{k}=\gamma$ and $\Delta=\cos(\gamma)$.  The
constant $\gamma$ is fixed by requiring that Eq.~(\ref{D[ro0]=0}) is
also satisfied, which amounts to meeting the boundary conditions
$\varphi_{1}=0$ and $\varphi_{N}=\Phi$.  We conclude that, for any
twisting angle $\Phi$, we have a factorized state which satisfies the
Zeno regime conditions (\ref{D[ro0]=0}) and (\ref{SecularConditions0})
only when the anisotropy assumes the values
\begin{align}
  \Delta(\Phi,m) = \cos\left( (\Phi+2 \pi m)/(N-1) \right),
  \label{DeltaForPureStates}%
\end{align}
with $m=0,1,\dots,N-2$.  This solution represents an equidistant
twisting of the polarization vector in the $XY$ plane along the chain
with winding number $m$; see Fig.~\ref{FigSpinsXY} for an
illustration.  For a fixed twisting angle $\Phi$ and in the limit
$N\rightarrow \infty$, the set of the
anisotropies~(\ref{DeltaForPureStates}) becomes dense in the interval
$[-1,1]$.  While the pure states that we obtain, are, by construction,
dark states of the quantum jump operators, $L_\alpha | \Psi
\rangle=0$, $\alpha=1,2$, they are not eigenstates of the Hamiltonian,
$H | \Psi \rangle \neq \lambda | \Psi \rangle$.

Since Eqs.~(\ref{D[ro0]=0}) and (\ref{SecularConditions0}) satisfied
by our $XY$-twisting solution are just necessary (but not sufficient)
conditions for the NESS in the Zeno limit to be pure, one needs an
independent check of the pureness of the found solution.  A
straightforward analytic computation for small system sizes $N\leq 7$
reveals that indeed the found NESS in the Zeno limit becomes pure
exactly for the anisotropies (\ref{DeltaForPureStates}), with two
exceptions, namely, $\Phi=0$ and $\Delta=0$; see Appendix
\ref{SM3}. Moreover, we find that no other pure states in the Zeno
limit exist. Thus, all solutions of Eqs.~(\ref{D[ro0]=0})
and (\ref{SecularConditions}) for real-valued $\Delta$ are given by
the factorized states (\ref{PsiFactorized}) with anisotropy
(\ref{DeltaForPureStates}).

\begin{figure}[ptbh]
  \begin{center}
    \includegraphics[width=0.99\columnwidth]{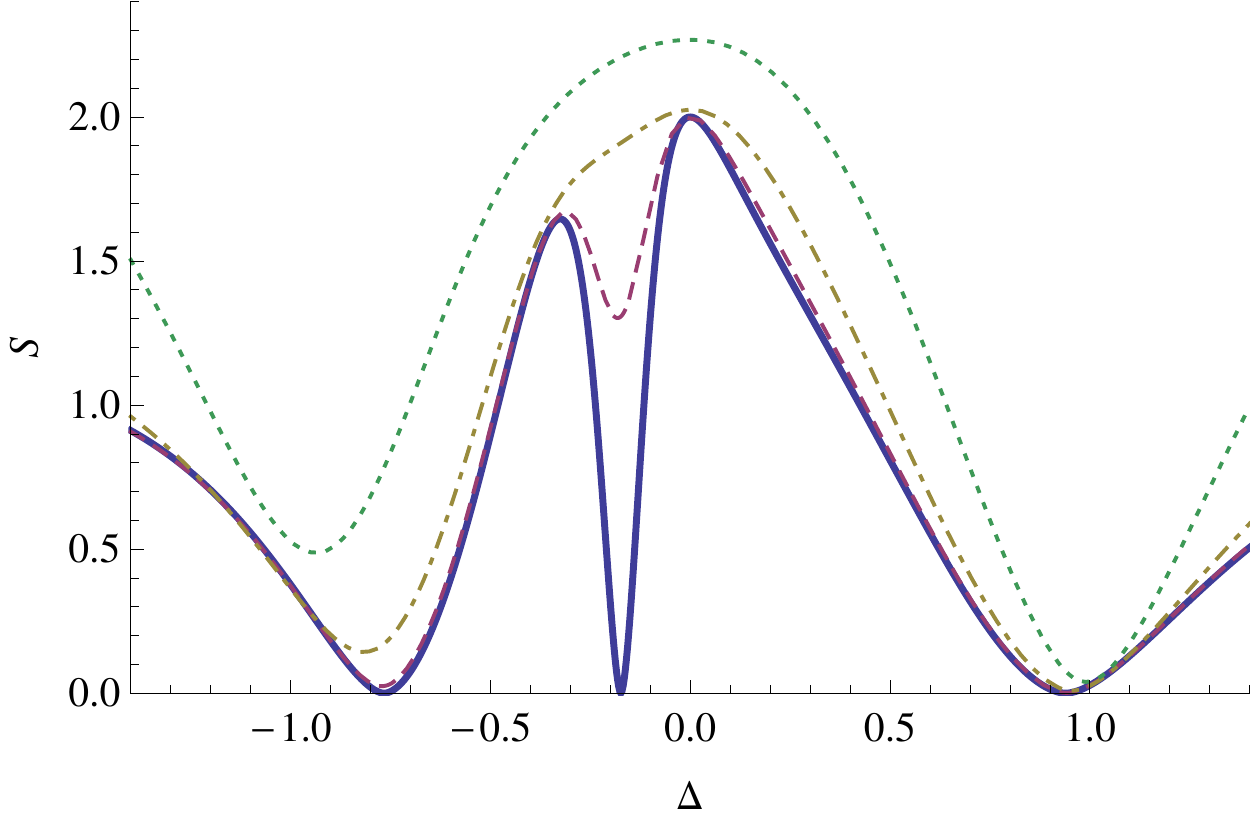}
  \end{center}
  \caption{(Color online) The von Neumann entropy $S=-\tr(\rho \log_2 \rho
    )$ for $\rho=\rho_\mathrm{NESS}$ versus the $Z$-axis anisotropy
    $\Delta$, in a driven $XXZ$ chain with boundary twisting in the
    $XY$ plane.  The dotted, dot-dashed and dashed lines correspond to
    finite dissipative couplings, $\Gamma=2,5,15$, respectively, while
    the thick line is the Zeno limit.  Parameters: $N=4$,
    $\Phi=\pi/3$.}
  \label{entropy_XY}%
\end{figure}

In Fig.~\ref{entropy_XY} we show the von Neumann entropy
$S=-\tr(\rho_\mathrm{NESS}\log_2 \rho_\mathrm{NESS})$ versus the
anisotropy $\Delta$, in the Zeno limit and for finite $\Gamma$,
obtained numerically for a system of four sites.  In the Zeno limit,
the NESS becomes pure, i.e., $S=0$, only at the points predicted by
Eq.~(\ref{DeltaForPureStates}). For finite $\Gamma$, the NESS is
always mixed.  However, at the points (\ref{DeltaForPureStates}), and
for $\Gamma$ finite but larger and larger, $\rho_\mathrm{NESS}$
approaches the respective pure states arbitrarily closely.

\begin{figure}[ptbh]
  \begin{center}
    \includegraphics[width=0.99\columnwidth]{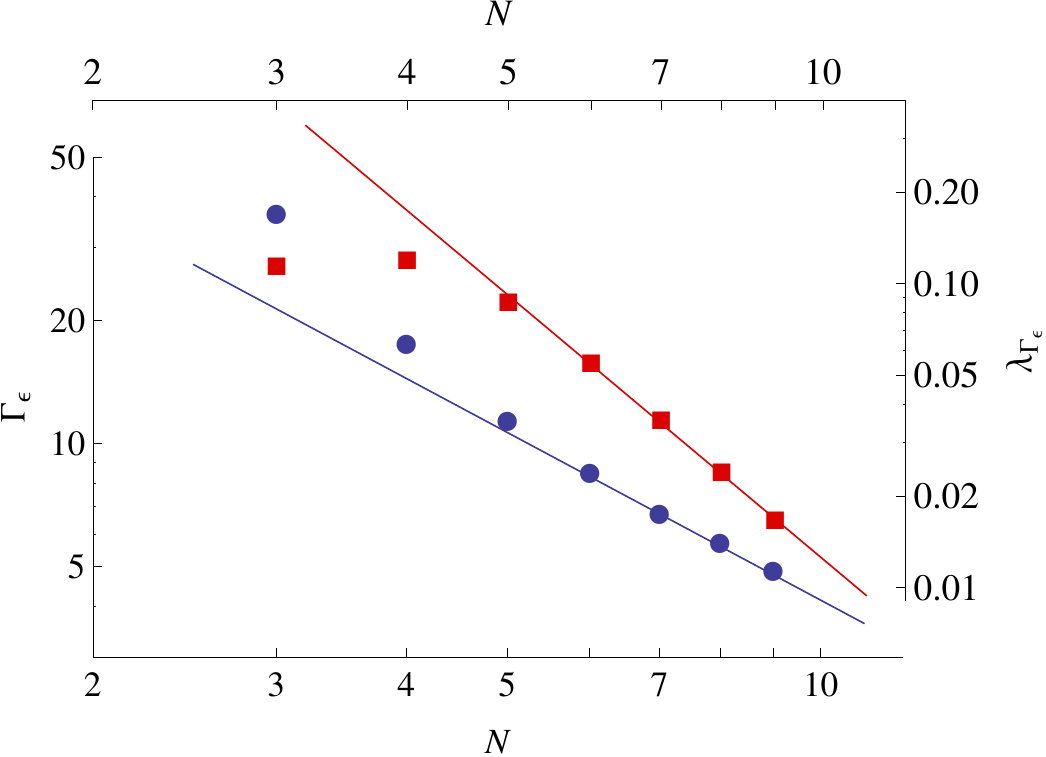}
  \end{center}
  \caption{(Color online) Minimal dissipation strength
    $\Gamma_\epsilon$ needed to reach a pure state with accuracy
    $\epsilon=10^{-3}$ versus system size $N$ (dots, left-bottom
    axes). The solid line is the fit $\Gamma_\epsilon = 94.8
    N^{-1.36}$.  Gap $\lambda_{\Gamma_\epsilon}$ of the Liouvillian
    spectrum at $\Gamma=\Gamma_\epsilon$ versus system size $N$
    (squares, right-top axes).  The solid line is the fit
    $\lambda_{\Gamma_\epsilon}=9.42 N^{-2.88}$.  Parameters:
    $\Phi=\pi/3$, $\Delta= \cos(\Phi/(N-1))$.}
  \label{GammaGap}%
\end{figure}

To find out if the pure states found in the Zeno regime are
experimentally accessible, we have numerically calculated the minimal
dissipation strength $\Gamma_\epsilon(m,N,\Phi)$ required to reach a
pure NESS within a given tolerance $\epsilon$, and the relaxation time
needed to establish the NESS, namely, the inverse gap
$\lambda(\Gamma_\epsilon)^{-1}$ of the spectrum of the Liouvillian
$\mathcal{L}[\cdot] = -i\left[ H,\cdot\right] +\Gamma
\mathcal{D}[\cdot]$ at $\Gamma=\Gamma_\epsilon$. In practice, we
define $\Gamma_\epsilon$ as the dissipation strength at which the von
Neumann entropy of the corresponding NESS becomes equal to $\epsilon$;
see Appendix \ref{SM4} for details.  Most remarkably, we find, on the base of a
study of small-size systems ($N\leq 9$), that the optimal (minimized
among all the winding numbers $m$ \cite{NotaCurrent})
$\Gamma_\epsilon$ \emph{decreases} with $N$, making the effective
``Zeno regime'' more and more accessible as the system size increases;
see Fig.~\ref{GammaGap}.  This somewhat counter-intuitive property
follows from the fact that for longer chains, it becomes easier to
freeze the boundary spins, i.e. to suppress their fluctuations, so
that the effective Zeno regime is reached earlier.  In compensation,
the corresponding relaxation time increases with $N$; see
Fig.~\ref{GammaGap}. However, this increase is only polynomial, in
accordance with the general observation made in
\cite{Znidaric2015Gaps}.

\subsection{Boundary twisting in the $XZ$ plane}  
\begin{figure}[ptbh]
  \begin{center}
    \includegraphics[width=0.3\columnwidth]{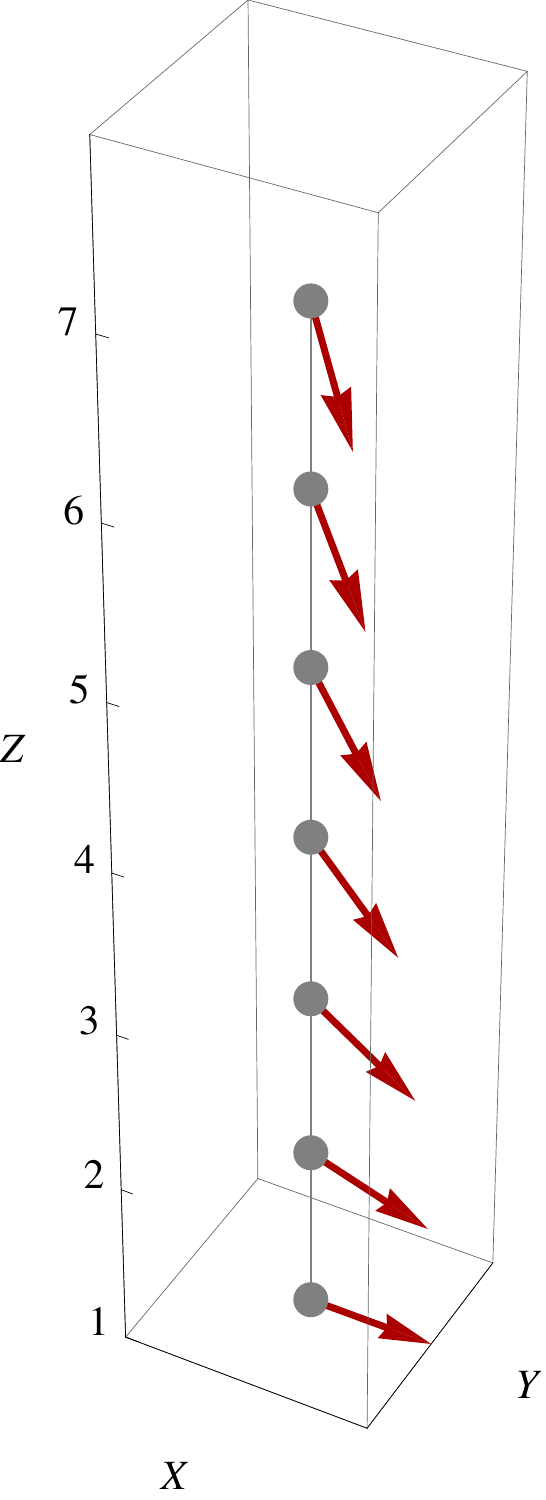}
    \includegraphics[width=0.3\columnwidth]{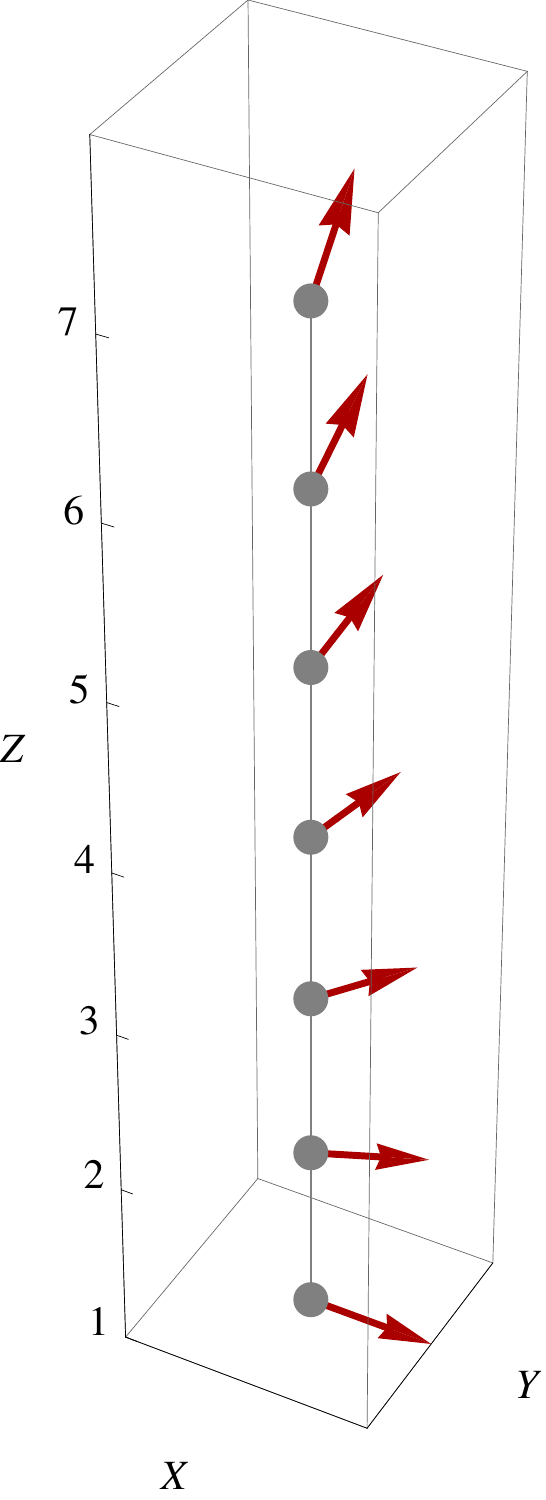}
  \end{center}
  \caption{ (Color online) As in Fig.~\ref{FigSpinsXY} for boundary
    twisting in the $XZ$ plane and in correspondence to solutions with
    decreasing (left) and increasing (right) orbital angles.
    Parameters: $N=7$, $\theta_L=\pi/2$, $\theta_R=\pi/10$ (left),
    $\theta_R=\pi-\pi/10$ (right).}
  \label{FigSpinsXZ}%
\end{figure}

Next we orient the boundary polarization in the $XZ$ plane, ${\vec
  l}_L = (\sin\theta_L,0,\cos\theta_L)$, ${\vec l}_R =
(\sin\theta_R,0,\cos\theta_R)$. As before, we first solve
Eq.~(\ref{GeneralizedDivergenceCondition}), now taking $| \psi_k
\rangle$ in the form
\begin{align}
  | \psi_k \rangle=
  \left( \begin{array}{l}
      \cos \frac {\theta_k}{2}
      \\
      \sin\frac {\theta_k}{2}
    \end{array}
  \right),
\end{align}
which corresponds to a local spin polarization ${\vec
  l}_k=(\sin\theta_k,0,\cos\theta_k)$, and then restrict the found
solution to meet Eqs.~(\ref{D[ro0]=0}) and (\ref{SecularConditions0}).
As detailed in Appendix \ref{SM5}, we have a NESS which can be of two kinds,
corresponding to the orbital angles $\theta_{k}$ monotonically
decreasing or increasing in the interval $]0,\pi[$; see
Fig.~\ref{FigSpinsXZ}.  We never have a pure NESS with
$\theta_{k}=0,\pi$ unless in the thermodynamic limit $N \rightarrow
\infty$.  For finite-size systems and given boundary polarizations in
the $XZ$ plane, i.e., $\theta_L$, $\theta_R$, the NESS becomes pure in
the Zeno limit and only for the anisotropy value,
\begin{align}
  \Delta (\theta_L,\theta_R)= &\frac{1}{2} \left[ \tan(\theta_{R}/2)/
    \tan(\theta_L/2) \right]^{1/(N-1)} \nonumber \\ &+ \frac{1}{2}
  \left[ \tan(\theta_{L}/2)/ \tan(\theta_R/2) \right]^{1/(N-1)}.
  \label{DeltaForPureStatesXZ}%
\end{align}

\begin{figure}[ptbh]
  \begin{center}
    \includegraphics[width=0.99\columnwidth]{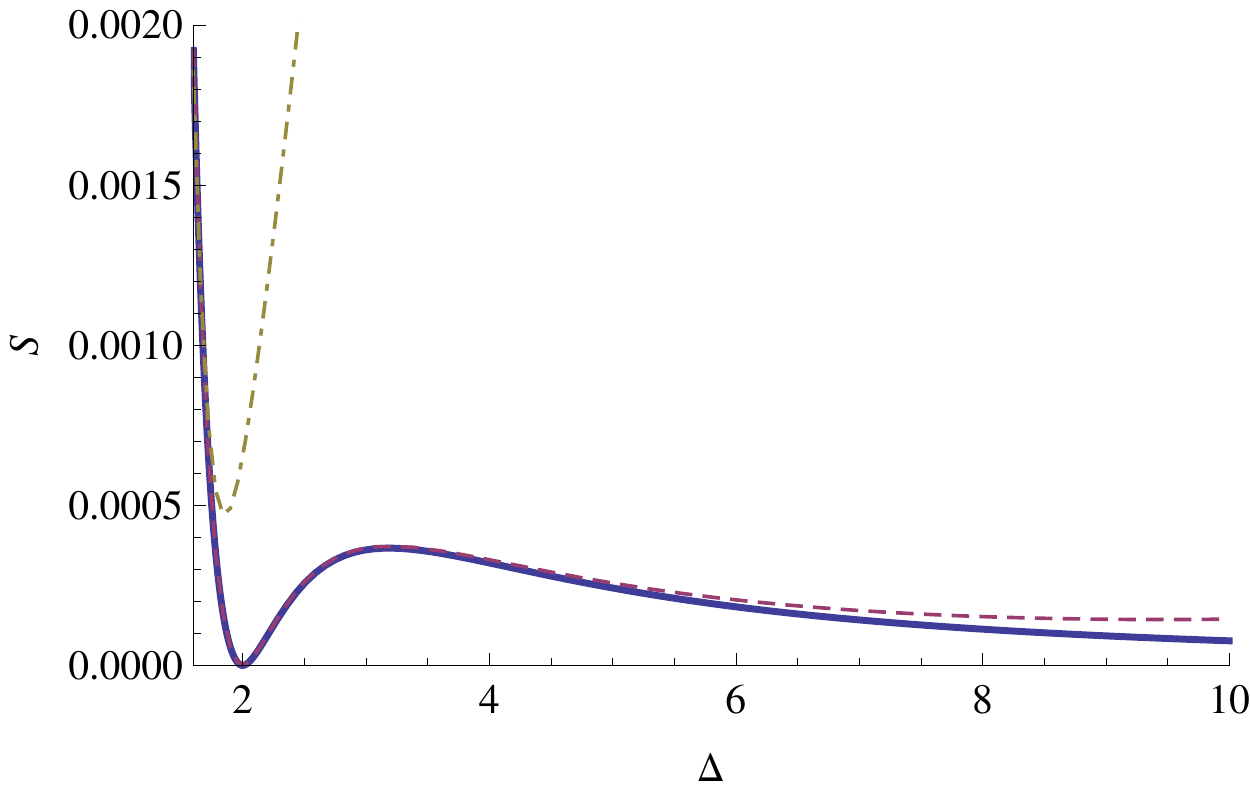}
  \end{center}
  \caption{ (Color online) The von Neumann entropy $S=-\tr(\rho \log_2
    \rho )$ for $\rho=\rho_\mathrm{NESS}$ versus the $Z$-axis
    anisotropy $\Delta$, in a driven $XXZ$ chain with boundary
    twisting in the $XZ$ plane.  Parameters: $N=4$, $\theta_L=\pi/2$,
    $\theta_R=\arctan(1/(15\sqrt{3}))$. The NESS is pure for a single
    value of the anisotropy, $\Delta=2$, given by
    Eq.~(\ref{DeltaForPureStatesXZ}).  The dot-dashed and dashed lines
    are obtained for finite dissipative couplings, $\Gamma=30,370$,
    respectively, while the thick line is the Zeno limit. }
  \label{entropy_XZ}%
\end{figure}

In Fig.~\ref{entropy_XZ}, we show the dependence of the von Neumann
entropy $S=-\tr(\rho_\mathrm{NESS} \log_2 \rho_\mathrm{NESS})$,
obtained by numerically evaluating the NESS in a system of four sites
for different values of the anisotropy. In the Zeno limit, the NESS
becomes pure, i.e., $S=0$, at the point predicted by
Eq.~(\ref{DeltaForPureStatesXZ}).

\section{Conclusions}
To summarize, we have formulated a criterion for a nonequilibrium
steady state of an open quantum system to be pure, in the Zeno limit,
i.e., for asymptotically large dissipative coupling.  The criterion is
specified by Eqs.~(\ref{D[ro0]=0}) and
(\ref{SecularConditions}). Zeno-limit pure states are not reachable,
in a strict mathematical sense, for any finite dissipative coupling.
However, by applying a finite but large enough dissipative coupling,
one can generate pure NESSs with arbitrary precision.

Using our criterion, in the Zeno regime we find two families of pure
NESSs for the driven quantum $XXZ$ spin chain with boundary twisting
in the $XY$ or $XZ$ plane, for values of the $Z$-axis anisotropy given
by Eqs.~(\ref{DeltaForPureStates}) and (\ref{DeltaForPureStatesXZ}),
respectively.  The criterion can be straightforwardly applied to
generate pure steady states in other nonequilibrium quantum systems.

Our approach opens an interesting perspective in dissipative
engineering of pure states.  If, for given resources, preparing an
exact pure steady state at finite dissipative strength is impossible
(as it happens in our example of driven $XXZ$ chain), it may still be
possible to generate a pure state in the Zeno limit. In practice, this
means that at finite dissipation, a slightly mixed state will be
produced, which, however, becomes infinitesimally close to a pure
state as the dissipation is increased. The effective coupling needed
to reach the ``Zeno regime'' depends on the chosen measure of pureness
and the required precision and must be estimated in each case
separately. In the example of the driven $XXZ$ model considered here,
the effective Zeno regime is reached at very moderate dissipative
couplings.

\begin{acknowledgments}
  VP thanks the Dipartimento di Fisica of Sapienza Universit\`a di
  Roma for hospitality and the Istituto Nazionale di Fisica Nucleare,
  Sezione di Roma 1, for partial support. VP also thanks
  M.~\v{Z}nidari\v{c}, G.~Sch\"utz and C.~Kollath for
  discussions. Financial support by the Deutsche
  Forschungsgemeinschaft is gratefully acknowledged.
\end{acknowledgments}

\appendix

\section{Inverse of the Lindblad dissipator and secular conditions}
\label{SM1}%
The Lindblad operators $L_1 \equiv \mathcal{L}_L$, $L_2 \equiv
\mathcal{L}_R$ have the form
\begin{align*}
  L_1 =&\
  [(\cos\theta_{L}\cos\varphi_{L})\sigma_{1}^{x}+(\cos\theta_{L}%
  \sin\varphi_{L})\sigma_{1}^{y} \nonumber\\ &
  -(\sin\theta_{L})\sigma_{1}^{z} +i\sigma_{1}^{x}(-\sin\varphi_{L})
  +i\sigma_{1}^{y}(\cos\varphi_{L})]/2,
%  \label{L_Left}%
  \\
  L_2 =&\
  [(\cos\theta_{R}\cos\varphi_{R})\sigma_{N}^{x}+(\cos\theta_{R}%
  \sin\varphi_{R})\sigma_{N}^{y} \nonumber\\ &
  -(\sin\theta_{R})\sigma_{N}^{z} +i\sigma_{N}^{x}(-\sin\varphi_{R})
  +i\sigma_{N}^{y}(\cos\varphi_{R})]/2.
%  \label{L_Right}%
\end{align*}
The dissipator, $\mathcal{D}=\mathcal{D}_L+\mathcal{D}_R$, is the sum
of the left and right dissipators
\begin{align*}
  \mathcal{D}_{L} [\cdot] = L_1 \cdot L_1^\dag - \frac{1}{2}\{L_1^\dag
  L_1,\cdot\},
  \\
  \mathcal{D}_{R} [\cdot] = L_2 \cdot L_2^\dag - \frac{1}{2}\{L_2^\dag
  L_2,\cdot\},
\end{align*}
which are linear super-operators acting locally on a single qubit. The
eigenbasis $\{\phi_{R}^{\alpha}\}_{\alpha=1}^{4}$ of the eigenproblem
$\mathcal{D}_{R}[\phi_{R}^{\alpha}]=\lambda_{\alpha}\phi_{R}^{\alpha}$
is
\begin{align*}
  \{\phi_{R}^\alpha\} = \{& 2\rho_{R},2\rho_{R} -I, -\sin\varphi_{R}
  \sigma^{x} + \cos\varphi_{R} \sigma^{y}, \nonumber \\ &\cos\theta_R
  (\cos\varphi_{R} \sigma^{x} + \sin\varphi_{R} \sigma^{y})- \sin
  \theta_R \sigma^{z}\}
\end{align*}
with the respective eigenvalues
\begin{align*}
  \{\lambda_{\alpha}\}=\{0,-1,-\frac{1}{2},-\frac{1}{2}\}.
\end{align*}
Here $I$ is a 2$\times$2 unit matrix, $\sigma^{x},\sigma
^{y},\sigma^{z}$ are the Pauli matrices, and $\rho_{R}$ is the
targeted spin orientation at the right boundary.  Analogously, the
eigenbasis and eigenvalues of the eigenproblem
$\mathcal{D}_{L}[\phi_{L}^{\beta}]=\mu_{\beta}\phi_{L}^{\beta}$ are
\begin{align*}
  \{\phi_{L}^\beta\}=\{&2\rho_{L},2\rho_{L} -I, -\sin\varphi_{L}
  \sigma^{x} + \cos\varphi_{L} \sigma^{y}, \nonumber \\ & \cos\theta_L
  (\cos\varphi_{L} \sigma^{x} + \sin\varphi_{L} \sigma^{y})- \sin
  \theta_L \sigma^{z}\},
\end{align*}
\begin{align*}
  \{\mu_{\beta}\}=\{0,-1,-\frac{1}{2},-\frac{1}{2}\},
\end{align*}
where $\rho_{L}$ is the targeted spin orientation at the left
boundary. Since the bases $\{\phi_{R}^\alpha\}$ and
$\{\phi_{L}^\beta\}$ are complete, any matrix $\chi$ acting in the
appropriate Hilbert space can be expanded as
\begin{align}
  \chi=\sum\limits_{\alpha=1}^{4}\sum\limits_{\beta=1}^{4}\phi_{L}^{\beta}
  \otimes
  \chi_{\beta\alpha}\otimes\phi_{R}^{\alpha}. \label{ro_expansion}%
\end{align}
Indeed, let us introduce two complementary bases
\begin{align*}
  \{\psi_{L,R}^\alpha\}=\{ &I/2, \rho_{L,R} -I, (-\sin\varphi_{L,R}
  \sigma^{x} + \cos\varphi_{L,R} \sigma^{y})/2, \nonumber \\ &
  (\cos\theta_{L,R} (\cos\varphi_{L,R} \sigma^{x} + \sin\varphi_{L,R}
  \sigma^{y}) \nonumber \\ &\quad - \sin \theta_{L,R} \sigma^{z})/2\},
\end{align*}
trace-orthonormal to the $\{\phi_{L,R}^\alpha\}$, namely,
$\tr(\psi_{R}^{\gamma}\phi_{R}^{\alpha})=\delta_{\alpha\gamma}$ and
$\tr(\psi_{L}^{\gamma}\phi_{L}^{\beta})=\delta_{\beta\gamma}$. Then,
the coefficients of the expansion (\ref{ro_expansion}) are given by
\begin{align*}
  \chi_{\beta \alpha}=\trb((\psi_{L}^{\beta}\otimes
  I^{\otimes_{N-1}})F(I^{\otimes _{N-1}}\otimes\psi_{R}^{\alpha})),
\end{align*}
where $\trb$ denotes the trace taken with respect to the first- and the
last-spin spaces only.  On the other hand, in terms of the expansion
(\ref{ro_expansion}), the super-operator inverse $\mathcal{D}^{-1}= (
\mathcal{D}_{L}+\mathcal{D}_{R})^{-1}$ is simply
\begin{align*}
  (\mathcal{D}_{L}+\mathcal{D}_{R})^{-1}[\chi]=
  \sum\limits_{\alpha,\beta}\frac{1}{\lambda_{\alpha}+\mu_{\beta}}
  \phi_{L}^{\beta}\otimes \chi_{\beta\alpha}\otimes\phi_{R}^{\alpha}.
%  \label{LindbladInversion}%
\end{align*}
The above sum contains a singular term with $\alpha=\beta =1$, because
$\lambda_{1}+\mu_{1}=0.$ To eliminate this singularity, one must
require $\chi_{11}=\trb \chi=0$, which is equivalent to the secular
condition
\begin{align*}
  P_{\ker \mathcal{D} } (\chi) = 0,
%  \label{SecularCondition}%
\end{align*}
where $P_{\Omega}$ denotes the orthogonal projector on $\Omega$.

We conclude that the existence of
$\rho^{(k+1)}=\mathcal{D}^{-1}[i[H,\rho^{(k)}]]$ at order
$k=0,1,2,\dots$ is granted if and only if
\begin{align}
  \trb ([H,\rho^{(k)}]) = 0.
  \label{AlternativeSecularConditions}%
\end{align}

\section{Stationary states with boundary twisting in the $XY$ plane}
\label{SM2}%
Assuming $\rho^{(0)} = | \Psi \rangle \langle \Psi |$ in the
factorized form
\begin{align*}
  | \Psi \rangle = | \psi_1 \rangle \otimes | \psi_2 \rangle \otimes
  \dots | \psi_{N} \rangle, 
%\label{PsiFactorized}%
\end{align*}
with
\begin{align*}
  | \psi_k \rangle= \frac{1}{\sqrt{2}} \left(
    \begin{array}{c}
      e^{-i\frac{\varphi_k}{2}}
      \\
      e^{i\frac{\varphi_k}{2}}
    \end{array}
  \right),\qquad k=1,\dots,N,
\end{align*}
we look for a solution of the generalized divergence condition
\begin{align}
  h \left(|\psi_k \rangle \otimes |\psi_{k+1}\right \rangle) =&\ \mu_k
  |\psi_k\rangle \otimes |\psi_{k+1}\rangle \nonumber\\ &+ |U_k\rangle
  \otimes |\psi_{k+1}\rangle - |\psi_{k}\rangle \otimes |U_k\rangle,
  \label{GeneralizedDivergenceConditionApp}%
\end{align}
where $h$ is the local density of the $H_{XXZ}$ Hamiltonian,
\begin{align*}
  h = \frac{1}{2} \left( \sigma^{x}\otimes \sigma^{x}+
    \sigma^{y}\otimes \sigma^{y} + \Delta \left( \sigma^{z}\otimes
      \sigma^{z}-I \right) \right),
\end{align*}
and $|U_k\rangle$ is a local unknown vector,
\begin{align*}
  |U_k\rangle=\left(
    \begin{array}{c}
      u_k
      \\
      v_k
    \end{array}
  \right).
\end{align*}
Equation~(\ref{GeneralizedDivergenceConditionApp}) is an overdetermined
system of equations for $\mu_k, u_k,v_k$.  The system does not admit a
solution unless the $Z$-anisotropy parameter takes the value
$\Delta=\cos(\varphi_{k+1}-\varphi_k)$, which is possible only if the
difference between any two consecutive angles along the chain is kept
constant, $\varphi_{k+1} - \varphi_{k}=\gamma$.  In this case, we have
%\begin{subequations}
  \begin{align*}
    &\mu_k = 4\sin \frac{\varphi_{k+1}-\varphi_k}{4}\cos^2
    \frac{\varphi_{k+1}-\varphi_k}{4},
    \\
    &u_k = -i\sqrt{2} \sin \frac{\varphi_{k+1}-\varphi_k}{4} \cos
    \frac{\varphi_{k+1}-\varphi_k}{2} e^{-i
      \frac{\varphi_{k+1}+\varphi_k}{4}},
    \\
    &v_k = i\sqrt{2} \sin \frac{\varphi_{k+1}-\varphi_k}{4} \cos
    \frac{\varphi_{k+1}-\varphi_k}{2} e^{i
      \frac{\varphi_{k+1}+\varphi_k}{4}}.
  \end{align*}
%  \label{solXY}%
%\end{subequations}

From the above solution, we compute
%\begin{subequations}
  \begin{align*}
    R_k &= | U_k \rangle \langle \psi_k | - | \psi_k \rangle \langle
    U_k | = \frac{i}{2} \sigma^z\sin(\varphi_{k+1}-\varphi_{k}),
    \\
    \tilde{ R}_k &= | U_{k-1} \rangle \langle \psi_k | - | \psi_k
    \rangle \langle U_{k-1} | = \frac{i}{2}
    \sigma^z\sin(\varphi_{k}-\varphi_{k-1}).
  \end{align*}
  % \label{RnDefinition}%
%\end{subequations}
It is straightforward to check that the system of equations,
\begin{subequations}
  \begin{align}
    &\sum_{k=1}^{N-1} \left(\mu_k-\mu_k^* \right) = 0,
    \\
    &R_k = \tilde{R}_{k},\qquad k=1,\dots,N,
    \\
    &\tr R_1 =\tr \tilde{R}_N= 0,
  \end{align}
  \label{ZerothOrderConditionsForRApp}%
\end{subequations}
is then satisfied.  Since Eq.~(\ref{ZerothOrderConditionsForRApp}) has
been demonstrated to be equivalent to 
$P_{\ker \mathcal{D} } ([H,| \Psi \rangle \langle \Psi |]) = 0$,
we conclude that the found solution meets the necessary
condition of our Zeno regime pure NESS criterion.  To meet the other
condition, namely, $\mathcal{D}[| \Psi \rangle \langle \Psi |] =0$, we
just need to satisfy the boundary conditions $\varphi_{1}=0$ and
$\varphi_{N}=\Phi$.  This is accomplished by choosing 
\begin{align*}
 \gamma =\gamma(\Phi,m)=(\Phi+2 \pi m)/(N-1), 
\end{align*}
with $m=0,1,\dots,N-2$.

\section{Analytic calculation of the Zeno NESS for small sizes and
 boundary twisting in the $XY$ plane}
\label{SM3}%
Solving the secular conditions (\ref{SecularConditions}) for $k=0,1$,
we compute the analytic form of $\rho_{NESS}(N,\Delta,\Phi)$ for small
system sizes $N$, and calculate the pureness parameter,
\begin{align*}
  f(N,\Delta,\Phi) = \tr[ \rho_{NESS}(N,\Delta,\Phi)^2 ]-1.
\end{align*}
Note that the state $\rho_{NESS}$ is pure if and only if $f=0$.  We
obtain
\begin{align*}
  &f(3,\Delta,\Phi) = -\frac{\left(-T_2(\Delta) +\cos (\Phi
      )\right)^2} {2 \left(2 \Delta ^2+\cos (\Phi )+1\right)^2},
  \\
  &f(4,\Delta,\Phi) = -\left(-T_3(\Delta)+\cos (\Phi )\right)^2
  \frac{P_4}{D_4} ,
  \\
  &f(5,\Delta,\Phi) =-\left(-T_4(\Delta)+ \cos (\Phi )\right)^2
  \frac{P_5}{D_5} ,
  \\
  &\vdots
\end{align*}
where $T_n(\cos x)=\cos(nx)$ are Chebyshev polynomials of the first
kind,
\begin{align*}
  &T_2(x) =-1+x^2 ,
  \\
  &T_3(x) =-3x+4x^3 ,
  \\
  &T_5(x) =1 -8x^2 +8 x^4 ,
  \\
  &\vdots
\end{align*}
and, for instance,
\begin{align*}
  P_4 =& 16 \Delta ^6+56 \Delta ^4+4 \left(10 \Delta^2+3\right) \Delta
  \cos (\Phi ) \\ &+ 30 \Delta ^2+3 \cos (2\Phi )+3,
\end{align*}
\begin{align*}
  Q_4 =& 2 \left(16 \Delta ^6+12 \Delta ^4+4\left(5 \Delta^2+2\right)
    \Delta \cos (\Phi )\right.  \\ &+ \left. 14\Delta ^2+\cos (2 \Phi
    )+1\right)^2.
\end{align*}

Substituting $\Delta=\cos\gamma$ into the above expressions for $f$, we
obtain
\begin{align*}
  &f(3,\cos\gamma,\Phi) = -\frac{(\cos (2 \gamma )-\cos (\Phi ))^2}{2
    (\cos (2 \gamma )+\cos (\Phi )+2)^2},
  \\
  &f(4,\cos\gamma,\Phi) = -\left(\cos (3 \gamma )-\cos (\Phi
    )\right)^2 \frac{P_4}{D_4},
  \\
  &f(5,\cos\gamma,\Phi) = -\left(\cos (4 \gamma )-\cos (\Phi
    )\right)^2 \frac{P_5}{D_5}, \\ &\vdots
\end{align*}
Extrapolating for arbitrary $N$, we get $\cos ((N-1) \gamma ) = \cos
(\Phi )$ as a pure NESS condition, yielding
\begin{align*}
  \gamma(\Phi,m)=(\Phi+2\pi m)/(N-1), \qquad m=0,1,...,N-2,
\end{align*}
as well as
\begin{align}
  \Delta(\Phi,m)=\cos\left( \gamma(\Phi,m)\right)
  = \cos\left(  \frac{\Phi+2 \pi m }{N-1}  \right).
  \label{DeltaForPureStatesApp}%
\end{align}
Independently, we verify that the anisotropy values given by
Eq.~(\ref{DeltaForPureStatesApp}) exhaust the solutions of the equations
$f(N,\Delta,\Phi)=0$ for $\Delta$ being real.  
Points of non-analyticity of the functions
$f(N,\Delta,\Phi)$, e.g. $f(3,0,\pi)=0/0$, correspond to exceptions
and need to be analyzed separately.

\textit{Exception (a)} $\Phi=0$. This case corresponds to a full
boundary alignment, i.e. to the absence of a boundary gradient.  The
Zeno NESS is pure only for $\Delta= 1$ for $N$ even, which corresponds
to $m=0$ in Eq.~(\ref{DeltaForPureStatesApp}), and $\Delta= \pm 1$ for
$N$ odd, corresponding to $m=0,(N-1)/2$ in
Eq.~(\ref{DeltaForPureStatesApp}). The $\Delta= 1$ solution represents a
trivial factorized state with all spins polarized in the $X$
direction. This homogeneous state remains a NESS for any finite value
of $\Gamma$.

\textit{Exception (b)} $\Delta=0$. Whenever among the critical
anisotropy values (\ref{DeltaForPureStatesApp}), a free fermion point
$\Delta=0$ appears, the respective NESS at $\Delta=0$ is not a pure
state, but a fully mixed state, apart from the boundaries,
$\rho_{NESS} =\rho_L \otimes (I/2)^{\otimes_{N-2}}\otimes \rho_R$. The
peculiarity of this exception results from the fact that the Zeno
limit $\Gamma \rightarrow \infty$ and the free-fermion limit $\Delta
\rightarrow 0$ do not commute, with the reason being the existence of an
extra symmetry of the NESS at $\Delta \rightarrow 0$; see
\cite{2012XYtwist} for an elaborate treatment of a $\Phi=\pi/2$ case.

\section{Minimal dissipation strength $\Gamma_\epsilon(m,N,\Phi)$}
\label{SM4}%
\begin{figure}[ptbh]
  \begin{center}
    \includegraphics[width=0.99\columnwidth]{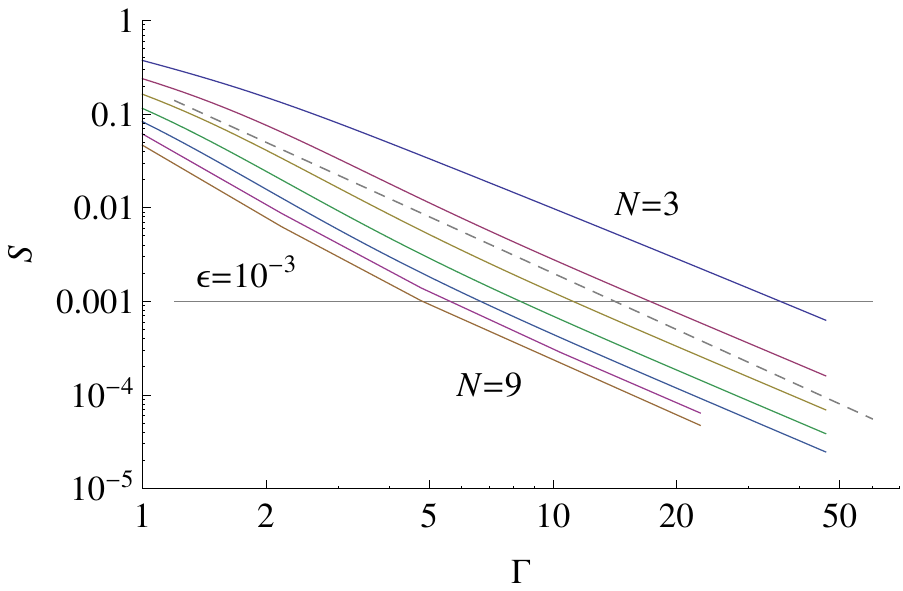}
  \end{center}
  \caption{(Color online) The von Neumann entropy
    $S(\Gamma)=-\tr[ \rho_\mathrm{NESS}(\Gamma) \log_2
    \rho_\mathrm{NESS}(\Gamma) ]$ corresponding to the NESS calculated
    for $m=0$ and $\Phi=\pi/3$ in the case of boundary twisting in the
    $XY$ plane and for system size $N$ from 3 (top) to 9 (bottom).
    The dashed line is the eye-guide $S\sim\Gamma^{-2}$ while the
    horizontal solid line is the tolerance value $S=10^{-3}$. }
  \label{S_vs_Gamma}%
\end{figure}

For system sizes $3\leq N \leq 9$, we have numerically calculated
$\rho_\mathrm{NESS}(\Gamma)$, namely the NESS of the Liouvillian
$-i[H,\cdot]+\Gamma \mathcal{D}[\cdot]$, where $H$ is the Hamiltonian
of the $XXZ$ model and $\mathcal{D}=\mathcal{D}_L+\mathcal{D}_R$ is
the dissipator described in Appendix~\ref{SM1}, for several finite
dissipation strengths $\Gamma$.  In the case of boundary twisting in
the $XY$ plane, the von Neumann entropy
$S(\Gamma)=-\tr(\rho_\mathrm{NESS}(\Gamma) \log_2
\rho_\mathrm{NESS}(\Gamma) )$ corresponding to the NESS obtained for
$m=0$ and $\Phi=\pi/3$ is plotted in Fig.~\ref{S_vs_Gamma} as a
function of $\Gamma$.  We see that, for any $N$, $S(\Gamma)$ decreases
monotonously to 0 by increasing $\Gamma$, approximately as
$\Gamma^{-2}$ for $\Gamma$ large.  As is natural, we define the minimal
dissipation strength $\Gamma_\epsilon(m,N,\Phi)$ required to reach a
pure NESS within a given tolerance $\epsilon$ as the unique solution
of
\begin{align*}
  S(\Gamma_\epsilon)=-\tr\left[ \rho_{NESS}(\Gamma_\epsilon) \log_2
  \rho_{NESS}(\Gamma_\epsilon) \right] =\epsilon,
\end{align*}
which is plotted in Fig.~3 of the main text.

\section{Stationary states with boundary twisting in the $XZ$ plane}
\label{SM5}%
Here, we solve Eq.~(\ref{GeneralizedDivergenceCondition}) taking $|
\psi_k \rangle$ in the form
\begin{align*}
  | \psi_k \rangle=
  \left( \begin{array}{l}
      \cos \frac {\theta_k}{2}
      \\
      \sin\frac {\theta_k}{2}
    \end{array}
  \right).
\end{align*}
It is convenient to denote
\begin{align*}
  \kappa_k=\left( \tan(\theta_{k+1}/2)  \right) /
  \left( \tan(\theta_k/2)  \right).
%  \label{kappa(k)}%
\end{align*}
We find that if the $Z$-anisotropy has the value
$2\Delta=\kappa_k+\kappa_k^{-1}$,
Eq.~(\ref{GeneralizedDivergenceCondition}) has the solution
%\begin{subequations}
  \begin{align*}
    &\mu_k = - \frac { \left( \cos \frac{\theta_{k} + 3
          \theta_{k+1}}{2} - \cos 3\frac{\theta_{k} +
          \theta_{k+1}}{2}\right)^2 } {\sin \theta_k \sin \theta_{k+1}
    },
    \\
    &u_k= \frac { \cos \frac{\theta_{k+1}}{2} - \cos
      \frac{\theta_k}{2} } {\sin \frac{\theta_k}{2} \sin
      \frac{\theta_{k+1}}{2} } \cos^2
    \frac{\theta_{k}+\theta_{k+1}}{4},
    \\
    &v_k= \frac { \sin \frac{\theta_{k+1}}{2} - \sin
      \frac{\theta_k}{2} } {\cos \frac{\theta_k}{2} \cos
      \frac{\theta_{k+1}}{2} } \sin^2
    \frac{\theta_{k}+\theta_{k+1}}{4}.
  \end{align*}
%  \label{solXZ}%
%\end{subequations}

From the above solution, we compute
%\begin{subequations}
  \begin{align*}
    &R_k = | U_k \rangle \langle \psi_k | - | \psi_k \rangle \langle
    U_k | = i b_k \sigma^y ,
    \\
    &\tilde{ R}_k = | U_{k-1} \rangle \langle \psi_k | - | \psi_k
    \rangle \langle U_{k-1} | = i \tilde{b}_k \sigma^y,
  \end{align*}
%\end{subequations}
with
%\begin{subequations}
  \begin{align*}
    b_k &= ( \cos\theta_{k+1}-\cos \theta_{k})/\sin \theta_{k+1},
    \\
    \tilde{b}_k &= (\cos \theta_{k}-\cos \theta_{k-1})/\sin
    \theta_{k-1}.
  \end{align*}
%\end{subequations}
Conditions (\ref{ZerothOrderConditionsForR}), i.e.,  $P_{\ker
  \mathcal{D} } ([H,| \Psi \rangle \langle \Psi |]) = 0$,
are thus satisfied if $\kappa_k+\kappa_k^{-1}=2\Delta$ and
$\tilde{b}_k=b_k$ for all $k$.  The latter condition after some
algebra gives
\begin{align}
  \kappa_{k}^{-1}+\kappa_{k-1}=\kappa_{k}+\kappa_{k-1}^{-1}.
  \label{kappaCondition}%
\end{align}
Notice that since $\kappa_{k}$ are real numbers, $|\Delta|>1$.  There
are two independent solutions of Eq.~(\ref{kappaCondition}), namely,
$\kappa_k = z_\pm$, where $z_\pm=\Delta \pm \sqrt{\Delta^2-1}$ are the
roots of the quadratic equation $z+1/z=\Delta$.  To meet the condition
$\mathcal{D}[| \Psi \rangle \langle \Psi |] =0$, we require
$\theta_{1}=\theta_{L}$ and $\theta_{N}=\theta_{R}$.

The solutions with $\kappa_k = z_\pm$ describe orbital angles
$\theta_{k}$ monotonically decreasing or increasing in the interval
$]0,\pi[$.  Note that we never have a pure NESS with
$\theta_{k}=0,\pi$, unless in the thermodynamic limit $N \rightarrow
\infty$.  In conclusion, for finite-size systems and given boundary
polarizations in the $XZ$ plane, we have one NESS in correspondence to
the anisotropy value,
\begin{align}
  \Delta (\theta_L,\theta_R)= &\frac{1}{2} \left[ \tan(\theta_{R}/2)/
    \tan(\theta_L/2) \right]^{\frac {1}{N-1}} \nonumber \\ &+
  \frac{1}{2} \left[ \tan(\theta_{L}/2)/ \tan(\theta_R/2)
  \right]^{\frac {1}{N-1}}.
  \label{DeltaForPureStatesXZApp}%
\end{align}


\begin{thebibliography}{99}
\bibitem{YamamotoRemark} The necessary and sufficient condition for a
  NESS to be pure can always be brought to the form
  (\ref{PureNESSsuffientCondition}) by a re-partitioning of $H$ and
  $L_\alpha$; see \cite{Yamamoto05,ZollerPRA08}.
\bibitem{NotaCurrent} We find that the optimal $m$ corresponds to the
  smallest steady-state current which, for $|\Phi|< \pi/2$, gives
  $m=0$ for $N$ even and $m=0,(N-1)/2$ for $N$ odd.
\end{thebibliography}
\end{document}